\documentstyle[12pt]{article}
\title{\bf Liquid-Droplet as a model for the rotation curve problem}
\author{F. Darabi \thanks{f.darabi@azaruniv.edu} \\
{\small Department of Physics, Azarbaijan Shahid Madani University, 53714-161 Tabriz, Iran}\\
{\small Research Institute for Astronomy and Astrophysics of Maragha (RIAAM), Maragha 55134-441, Iran}}
\begin{document}
\maketitle 
\begin{abstract}
The dynamics of large scale gravitational structures like galaxies, local groups and clusters is studied based on the so-called {\it Liquid-Droplet} model describing the saturation property of the nuclear force. Using the assumption that the gravitational force is also saturated over large scale structures, it is argued that the Newtonian gravitational potential may be replaced by an effective {\it Machian} gravitational potential. Application of this new potential at these large scale structures
may give the rotation curves in good agreement with observations. Then, the Virial theorem for this kind of gravitational interaction is developed and also the Tully-Fisher relation is obtained. A physical explanation is given for the so-called {\it constant} acceleration in the MOND as the
{\it effective} gravitational strength of these structures. Finally, a brief
argument is given for comparison with dark matter mode.
\\
Pacs: 98.62.Dm 
\end{abstract}

\newpage

\section{Introduction}

It is well known that classical Newtonian dynamics fails on galactic
scales. There is astronomical and cosmological evidence for a
discrepancy between the dynamically measured mass-to-light ratio of
any system and the minimum mass-to-light ratios that are compatible
with our understanding of stars, of galaxies, of groups and clusters
of galaxies, and of superclusters.  Observations on the rotation curves have turn out
that galaxies are not rotating in the same manner as the Solar
System. If the orbits of the stars are governed solely by
gravitational force, it was expected that stars at the outer edge of
the disc would have a much lower orbital velocity than those near
the middle. In fact, by the Virial theorem the total kinetic energy
should be half the total gravitational binding energy of the
galaxies. Experimentally, however, the total kinetic energy is found
to be much greater than predicted by the Virial theorem. Galactic
rotation curves, which illustrate the velocity of rotation versus
the distance from the galactic center, cannot be explained by only
the visible matter. This suggests that either a large portion of the
mass of galaxies was contained in the relatively dark galactic halo
or Newtonian dynamics does not apply universally.

The dark matter proposal is mostly referred to Zwicky \cite{Zwicky}
who gave the first empirical evidence for the existence of the
unknown type of matter that takes part in the galactic scale only by
its gravitational action. He found that the motion of the galaxies
of the clusters induced by the gravitational field of the cluster
can only be explained by the assumption of dark matter in addition
to the matter of the sum of the observed galaxies. Later, It was
demonstrated that dark matter is not only an exotic property of
clusters but can also be found in single galaxies to explain their
flat rotation curves. The dark matter problem has also been addressed through extended theories of gravity \cite{Dark}

The second proposal results in the modified Newtonian dynamics
(MOND), proposed by Milgrom, based on a modification of Newton's
second law of motion \cite{MOND}. This well known law states that an
object of mass $m$ subject to a force $F$ undergoes an acceleration
$a$ by the simple equation $F=ma$. However, it has never been
verified for extremely small accelerations which are happening at
the scale of galaxies. The modification proposed by Milgrom was the
following
\begin{equation}\label{mu(x)}
F=m\mu(\frac{a}{a_0})a,
\end{equation}
\begin{equation}
\mu(x)=\left \{ \begin{array}{ll} 1 \:\: \mbox{if}\:\: x\gg 1
\\
x \:\: \mbox{if}\:\: \|x|\ll 1,
\end{array}\right.
\end{equation}
where $a_0=1.2\times10^{-10} ms^{-2}$ is a proposed new constant.
The acceleration $a$ is usually much greater than $a_0$ for all
physical effects in everyday life, therefore $\mu(a/a_0)$=1 and
$F=ma$ as usual. However, at the galactic scale outside the central bulge
where $a \sim a_0$ we have the modified dynamics $F=m(\frac{a^2}{a_0})$. Using this new law of dynamics for the gravitational force $F=\frac{GmM}{r^2}$
we obtain
\begin{equation}
\frac{GM}{r^2}=(\frac{a^2}{a_0}),
\end{equation}
which results in the constant rotational velocity 
\begin{equation}\label{v}
v^2=\sqrt{GMa_0}.
\end{equation}
Another interesting model in this direction has been recently
proposed by Sanders. In this model, it is assumed that gravitational
attraction force becomes more like $1/r$ beyond some galactic
scale\cite{Sanders}. A test particle at a distance $r$ from a large
mass $M$ is subject to the acceleration
\begin{equation}
a = \frac{GM}{r^2} g(r/r_0),
\end{equation}
where $G$ is the Newtonian constant, $r_0$ is of the order of the
sizes of galaxies and $g(r/r_0)$ is a function with the asymptotic
behavior
\begin{equation}\label{g(r)}
g(r/r_0)=\left \{ \begin{array}{ll} 1 \:\: \mbox{if}\:\: r\gg r_0
\\
r/r_0 \:\: \mbox{if}\:\: r\ll r_0.
\end{array}\right.
\end{equation}

Dark matter as the manifestation of Mach principle has also been
considered as one of the solutions for the dark matter problem.
According to Mach principle \cite{Mach} the distant mass
distribution of the universe has been considered as being
responsible for generating the local inertial properties of the
close material bodies. Borzeszkowski and Treder have shown that the
dark matter problem may be solved by a theory of Einstein-Mayer type
\cite{Treder}. The field equations of this gravitational theory
contain hidden matter terms, where the existence of hidden matter is
inferred solely from its gravitational effects. In the
nonrelativistic mechanical approximation, the field equations
provide an inertia-free mechanics where the inertial mass of a body
is induced by the gravitational action of the cosmic masses. From
the Newtonian point of view, this mechanics shows that the effective
gravitational mass of astrophysical objects depends on $r$ such that
one expects the existence of new type of matter, the so-called dark
matter. In what follows we will introduce a new model for the dynamical structure of a galaxy, based on a model for the structure of the nucleus, as an alternative to dark matter and MOND to explain the rotation curve.  

\section{Saturation of the nuclear force}

In nuclear physics, Weizs\"acker's formula or the semi-empirical mass formula (SEMF) is used to approximate the mass and various features of atomic nucleus
\cite{Weizs}. The theoretical basis of this formula is the liquid drop model. This model in nuclear physics, first proposed by George Gamow and developed by Niels Bohr and John Archibald Wheeler, treats the nucleus as a drop of incompressible nuclear fluid which is made of nucleons (protons and neutrons) binding together by the strong nuclear force. It predicts the binding energy of a nucleus in terms of the numbers of protons and neutrons it contains. This statement
manifests in the semi-empirical mass formula which has five terms on its right hand side. These correspond to i) volume binding energy of all the nucleons inside the
nucleus, ii) surface binding energy of all the nucleons on the surface of
nucleus, iii) the electrostatic mutual repulsion of the protons, iv) an asymmetry term and v) a pairing term (both derivable from quantum mechanical considerations).

The first two terms concern about the {\it attractive} strong nuclear force.
We shall focus on these two terms and leave the other terms irrelevant for
our purposes. Assuming an approximately constant density for the nucleus one can calculate the nuclear radius by using that density as if the nucleus were a drop of a uniform liquid. The liquid droplet model of the nucleus takes into account the fact that the forces on the nucleons on the surface is different from those associated with the interior ones. This is something similar to taking into account of surface tension as a contributor to the energy of a tiny liquid drop. 

In the case of volume binding energy, when an assembly of nucleons of the same size is packed together into the smallest volume, each interior nucleon has a certain number of other nucleons in close contact with it. Actually, the number of pairs that can be taken from $A$ particles is $A(A-1)/2$, so one may expect a term for binding energy proportional to $A^2$ interacting pairs. However, the strong force has a very limited range, and a given nucleon may only interact strongly with its nearest neighbors. In other words, the nuclear force is saturated and each nucleon contributes an almost constant energy to the binding of the nucleus. Therefore, the number of pairs of particles that effectively interact is roughly proportional to $A$ and not $A^2$, and the nuclear binding energy is proportional to the volume of the nucleus. The volume term suggests that each nucleon interacts with a constant number of surrounding
nucleons, independent of $A$. 

In the case of surface binding energy however, each nucleon at the surface of a nucleus interacts with fewer other (interior and surface) nucleons than one in the interior of the nucleus. Therefore, the binding energy associated with each surface nucleon is less than that of the interior one. In fact, the surface term makes correction to the volume term by its negative contribution as follows  
\begin{equation}\label{1}
B=a_V A-a_S A^{2/3},
\end{equation}
where $B$ accounts for the binding energy and $a_V, a_S$ are some
appropriate constants with the same order of magnitude. One may focus more
on the volume term to make maximum use of its physical justification. The
physics behind this term in the case of a real liquid-droplet is as follows: To evaporate one liquid-droplet we have to use a definite heat energy as $Q_v M_m A$ where $Q_v$ accounts for the {\it vaporization latent heat}, $M_m$ is the mass of each molecule and $A$ is the total number of molecules in the
droplet. This heat energy is necessary to overcome all the attractive interactions
between the molecules. Hence, it is exactly equal to the binding energy of the droplet. This justifies the first term as 
\begin{equation}\label{2}
B_V=Q_v M_m A=a_V A
\end{equation}
which results in the identification $a_V=Q_v M_m$. Therefore, we come to the conclusion that the binding energy per each molecule, namely $(B/A)$, corresponding to the volume term  is independent of the total number of molecules $A$. This has a simple reason: the number of nearest neighbours of each interior
molecule is independent of the total size of the system (droplet), then $(B/A)$ becomes independent of $A$. {\it This is characteristic feature of every system in which the range of interaction between the constituent particles is effectively short in comparison with the size of the system}. Of course, this feature undergoes a surface correction due to the finite size of the system according to (\ref{1}).
Moreover, if we consider the repulsive coulomb force between the protons,
then its negative contribution proportional to $Z^2$ ($Z$ is the atomic
number) affects the ratio $(B/A)$ in a more complicate way. One may add
also the contributions of an asymmetry term and a pairing term. In general,
all of these contributions to the volume term except for the pairing term
are negative. 

The result is that the curve of $(B/A)$ with respect to $A$
is almost flat except for the small and large values of $A$. The decrease for small values of $A$, is due to the fact that the
full binding of each nucleon is accomplished when that nucleon is fully encircled
by other neighbor nucleons. However, this does not apply for the surface
nucleons; then for light nucleus having larger portion of surface nucleons
the binding energy for each nucleon in the nucleus is decreased. On the other
hand, the decrease for large values of $A$ results from the coulomb repulsive
force between each protons which is proportional to $Z^2$ and hence increases
more rapidly than $A$, resulting in a decrease of binding energy per each
nucleon. It is common to associate the almost flat portion of the curve with
the saturation property of the nuclear force which itself is emerged due
to the short range of this force. If this force would not be saturated, then each nucleon would interact with all other nucleons with a term proportional to the size of the nucleus or $A^2$. The more large size of the nucleus, the more stabilized system, a result which is not favored by the nature. Hence, it seems the saturation property or short range of the nuclear force is related to the fact that nature avoids the structures with unlimited large size. 

\section{Modified gravitational potential }

To model the large scale gravitational structures on the basis of Liquid-Droplet model we assume that gravity beyond the length scales larger than the scale of its inner component, namely over the halo, may be modified to represent the saturation property. Hence, in order to establish a gravity force which is saturated within the halo, we should propose a new kind of gravitational potential, different from Newtonian one. Motivated by the Liquid droplet model, we propose a gravitational potential energy as
\begin{equation}\label{3}
U(r)=k(r)N\left(-G\frac{m^2}{R}\right).
\end{equation}
Here $U(r)$ is the gravitational potential energy of mass $m$ located at
$r$ in any stable large scale gravitational structure containing $N$ gravitational constituents of average mass $m$, $R$ is the characteristic size of the structure, $G$ is the gravitational constant, and $k(r)$ is a function which together with $N$ determines the effective number of mutual Machian interactions $G{m^2}/{R}$ between the mass $m$ located at $r$ and any other mass $m$ in the structure. 
We know that the mass is a positive quantity and it is not possible to screen the gravitational effects. Therefore, we assume that in any two-body process in a given structure the extra influence of all other matter distributed over the characteristic size of the structure should be taken into account in a Machian way. 
Accordingly, we can replace the whole gravitational effect of all effective
interactions at the position of a constituent of mass $m$ by the effect of a spherical shell of the {\it effective} mass $M$ and the effective radius $R$. This shell acts like a gravitational Faraday cage inside of which a gravitational potential exists 
\begin{equation}\label{3'}
\Phi=-\frac{GM}{R}.
\end{equation}
Comparing this with the gravitational potential energy (\ref{3}) through $U=m \Phi$, we obtain the effective mass $M$ as an {\it active} gravitational mass
\begin{equation}\label{4}
M(r)=k(r)Nm.
\end{equation}
This is a key feature of the present model in which the active gravitational
mass is different from the additive mass of the galaxy, namely $Nm$.  
The need for this function is justified because the gravitational potential
(\ref{3}) without a local $r$-dependence will lead to a vanishing gravitational force. The very important function $k(r)$ certainly should depend on the local matter density distribution over the structure.

One may give different interpretations for the potential (\ref{3})
by rewriting it in the following forms
\begin{equation}\label{3''}
U(r)=-G\frac{(k(r)m)(Nm)}{R}.
\end{equation}
\begin{equation}\label{3'''}
U(r)=-G\frac{(m)k(r)(Nm)}{R}.
\end{equation}
\begin{equation}\label{3''''}
U(r)=-(G k(r))\frac{(m)(Nm)}{R}.
\end{equation}
The first potential describes the gravitational interaction between a local active (or Machian) mass $(k(r)m)$ and a global mass $(Nm)$, where the active mass describes a mass $m$ which is strengthened or weakened by the function $k(r)$. The second potential describes the gravitational interaction between a mass $m$ and a global mass $(Nm)$ through $k(r)$ which acts as a coupling
term strengthening or weakening the interaction. Finally, the third potential describes the gravitational interaction between a mass $m$ and a global mass $(Nm)$ through a modified gravitational coupling $G k(r)$ which is strengthened or weakened by $k(r)$. All these interpretations certify that $k(r)$ plays
the role of a {\it correlation function} between a mass $m$ (an star) located at $r$, and the surrounding global mass of the galaxy $(Nm)$ whose center of mass is located at the origin $r=0$. Therefore, we expect that the correlation function $k(r)$ is strengthened or weakened if the mass $m$ is located at dense or void places, respectively. 
\section{Rotation curve }

In this section, we study the rotation curve of a typical galaxy upon
the Machian gravitational potential. To this end, we focus on the important role of the correlation function $k(r)$ in the potential. It is important to note that we do not mean by $k(r)$ the density distribution of a galaxy, rather it is a correlation function which, in principle, can be extracted by a knowledge of the density distribution. For example, one may find the correlation function $k(r)$ by relating it to the Fourier space {\it matter power spectrum} of the mass distribution in the galaxy. But in practice, finding the actual correlation function $k(r)$ which can be derived directly from the density distribution is really a difficult task, especially for the halo component with a very low density distribution. To overcome this problem in our model, one may benefit an inverse procedure in that the knowledge about the observed rotation curve of a given galaxy can constrain at least the $r$-dependence of the correlation function $k(r)$ for that galaxy. In other words, by a numeric analysis one may constrain the appropriate $k(r)$ through the comparison of the {\it predicted} rotation curve given by
\begin{equation}
\frac{m v^2}{r}=-\frac{dk(r)}{dr}N\left(-G\frac{m^2}{R}\right),
\end{equation}
with the {\it observed} rotation curve\footnote{This is similar to constraining the galactic constants by use of the rotation curve \cite{HS}.}. By applying
this procedure for a large number of galaxies, perhaps one may succeed to
obtain an algorithm by which $k(r)$ can be derived from the density distribution. 

The idea that every galaxy with its characteristic
$k(r)$ will have its own rotation curve, different from other
ones especially in the halo, may be in agreement with the observations found
in \cite{SR}, \cite{HS}, where it is shown that the rotation curves
in the halo are not universal, rather some rotation curves are constant, some are increasing, some are declining, and some are exhibiting a true Keplerian decline. 
In other words, in the present model we have a characteristic correlation function $k(r)$ corresponding to each galaxy which is capable of producing constant, increasing, declining and even Keplerian rotation curves for the halo.

Let us examine this model to predict for example the flat rotation curves. If we assume the dense interior components of the galaxy to have roughly a disk shape with roughly constant density\footnote{Note that, the assumption of constant density may not be in complete agreement with observations, but even in dark matter model this assumption is used to justify the linear velocity profile of the rotation curve.}, then one expects that this component is
almost rigid and so the probability for observing a mass $m$ at the distance $r$ from the center, inside this rigid component, is proportional to the surface element, namely $r^2$. Hence, we may propose the following correlation
function
\begin{equation}\label{8}
k(r)=\alpha  r^2,
\end{equation}
where $\alpha$ is a dimensional constant. Note that this correlation function
strengthen the mass $m$, or equally results in strong active gravitational mass $M$. The potential (\ref{3}) is then written as 
\begin{equation}\label{9}
U(r)=-A r^2,
\end{equation}
where $A=\alpha{G Nm^2}/{R}$. By resorting to the Newton's second law we have
\begin{equation}\label{16}
\frac{m v^2}{r}=-\frac{dU}{dr}=2Ar,
\end{equation}
which gives the correct velocity profile for the interior component of galaxy
as
\begin{equation}\label{17}
v \propto r.
\end{equation}

For the halo, the density distribution of constituents is so low.  For example, in Milky way the halo distribution follows $r^{-3.5}$ which indicates that
the density of halo is almost ignorable in comparison to the inner dense
and rigid component of the galaxy. We know that the Milky way galaxy has {\it almost} a flat rotation curve. As discussed above, finding $k(r)$ analytically (corresponding to the density distribution $\sim r^{-3.5}$) which produces a rotation curve in exact agreement with that of Milky way is not easy at all. However, an oversimplified choice of $k(r)$ for galaxies like Milky way with almost flat rotation curve may be taken as a simple ansatz 
\begin{equation}\label{12}
k(r)=\ln{r},
\end{equation}   
for which we obtain
\begin{equation}\label{13}
U(r)=-B \ln{r},
\end{equation}
where $B=GNm^2/R$. 
The Newton's second law then becomes
\begin{equation}\label{18}
\frac{m v^2}{r}=-\frac{dU}{dr}=\frac{B}{r},
\end{equation}
which gives theoretically a flat rotation curve as
\begin{equation}\label{19}
v =\mbox{Const,}
\end{equation}
in {\it approximate} agreement with the {\it real} rotation curve of the Milky way. Of course, one may hope to produce the  {\it real} rotation curve of the Milky way provided that we know the exact form of correlation
function $k(r)$ (instead of  $\ln{r}$) for this galaxy.

To produce the increasing, declining and even Keplerian rotation curves,
we may try the ansatz  
\begin{equation}\label{8'}
k(r)=\alpha  r^n,
\end{equation}
and constrain the power $n$, respectively as
\begin{equation}\label{8''}
\left \{ \begin{array}{ll} 0<n<2\,\,\,\,\, \:\:\,\,\,\,\, \mbox{increasing},
\\
-1<n<0 \,\,\,\:\:\,\,\, \mbox{declining},
\\
n=-1 \,\,\,\:\: \,\,\,\,\,\,\,\,\,\,\,\,\,\mbox{Keplerian}.
\end{array}\right.
\end{equation}
Note that the above results, which ideally are in good agreement with observations,
come out due to our specific and simple assumptions about the forms of the correlation function $k(r)$, whereas in principle it may be a complicate
function of $r$ depending on the complexity of the density distribution or
mass power spectrum in different galaxies. However, it is reasonable to assume that the correlation function $k(r)$ for the dense and rigid inner component be almost the same for all galaxies, hence the velocity profile of this inner
component will be almost universal for most of the galaxies. But, for the halo the correlation function $k(r)$ may highly depend on the different characteristic features of each galaxy like density distribution,
shape of spiral arms, etc. Therefore, in this model it is expected that the flatness and slope of rotation curve for the halo may differ for different galaxies according to the diversity of $k(r)$ and its detailed characteristic features; a result which is in agreement with observations \cite{SR}, \cite{HS}.

\section{Virial theorem }

For a system of N point particles, the virial $G$ satisfies the following
equation \cite{Virial}
\begin{equation}\label{22}
\frac{dG}{dt}=2T+\sum_{i=1}^{N}{\bf F}_i \cdot {\bf r}_i,
\end{equation}
where $T$ is the total kinetic energy of the system, ${\bf F}_i$ is the net force on the $i^{th}$ particle at the position ${\bf r}_i$, and $G$ is defined
as 
\begin{equation}\label{23}
G=\sum_{i=1}^{N}{\bf P}_i \cdot {\bf r}_i.
\end{equation}
We have realized that in the Liquid-Droplet model, the total force on each particle $i$ is acted by neighbor particles. Moreover,
this number depends on the distance of that particle from the center of
galaxy. Therefore, the mutual interaction between all particles which is
usually considered in a gravitational system has now been replaced by this new type of limited interaction. If we assume the force to be derived from a potential energy, then we should accept that this potential is just a function of the distance of each particle from the center of galaxy $r_{i}$ rather than the relative distance $r_{ji}$ between the particles, namely ${\bf F}_{i}=
-{\bf \nabla}_{r_{i}}U$. Therefore, we have 
\begin{equation}\label{25}
\sum_{i=1}^{N}{\bf F}_i \cdot {\bf r}_i=-\sum_{i=1}^{N}
\frac{dU}{dr_{i}}r_{i}.
\end{equation} 
Hence, Eq.(\ref{22}) becomes
\begin{equation}\label{26}
\frac{dG}{dt}=2T-\sum_{i=1}^{N}
\frac{dU}{dr_{i}}r_{i}.
\end{equation}
For the galactic bulge with $U_{in}\sim r_i^2$ we obtain
\begin{equation}\label{27}
\frac{dG}{dt}=2T-2\sum_{i=1}^{N_{in}}
U_{in}(r_{i})=2T-2U_{in},
\end{equation}
where $U_{in}$ is the total potential energy of the system in the galactic bulge consisting of $N_{in}$ particles. On the other hand, outside the bulge
of the galaxy consisting of $N_{out}$ particles with $U_{out}=-B \ln{r_i}$, we have
\begin{equation}\label{28}
\frac{dG}{dt}=2T+\sum_{i=1}^{N_{out}}B
=2T+N_{out}B.
\end{equation}
The virial theorem states that the time average $\langle dG/dt\rangle$ vanishes,
hence we obtain
\begin{equation}\label{29}
\langle T \rangle_{in}=\langle U_T \rangle_{in},
\end{equation}
for the inside, and 
\begin{eqnarray}\label{30}
\langle T \rangle_{out}=-\frac{1}{2}N_{out}B ,
\end{eqnarray}
for the outside of the galactic bulge . Therefore, the total time average of the kinetic energy of the whole galaxy is obtained as
\begin{eqnarray}\label{31}
\langle T \rangle_{tot}&=&\langle U_T \rangle_{in}-\frac{1}{2}N_{out} B\\ \nonumber
&=&\frac{GNm^2}{R}\left(\sum_i^{N_{in}} {\langle k_{in}\rangle}-\frac{1}{2}N_{out} \right).
\end{eqnarray}
By interpreting $N_{out} B$ as $\langle U_T \rangle_{out}$, we obtain
\begin{eqnarray}\label{32}
\langle E \rangle_{in}=\langle T \rangle_{in}+\langle U_T \rangle_{in}=2\langle T \rangle_{in},
\end{eqnarray}
\begin{eqnarray}\label{33}
\langle E \rangle_{out}=\langle T \rangle_{out}+\langle U_T \rangle_{out}=-\langle T \rangle_{out},
\end{eqnarray}
\begin{eqnarray}\label{34}
\langle E \rangle_{tot}=2\langle T \rangle_{in}-\langle T \rangle_{out}.
\end{eqnarray}
In a stable gravitationally bound system we require $\langle E \rangle_{tot}<0$ which leads to
\begin{eqnarray}\label{35}
\langle T \rangle_{in}<\frac{1}{2}\langle T \rangle_{out}.
\end{eqnarray}
This is an interesting result which states that the total average kinetic energy in the galactic bulge is less than the half total average kinetic energy in the outside of galactic bulge.

\section{Tully-Fisher relation }

The Tully-Fisher relation is an empirical relationship between the intrinsic luminosity (proportional to the stellar mass) of a spiral galaxy and its amplitude of its rotation curve (which sets the total gravitational mass)
\cite{TF}. It has been turned out that on large scales most astronomical systems like
a spiral galaxy have much larger mass-to-light ratios than the central parts.
This fact may be explained by the relation (\ref{4}) as follows. In the present model, the gravitational mass is the active mass $M$. On the other hand,
the luminosity which is proportional to the stellar mass becomes proportional
to the additive mass $Nm$.  Therefore, the mass-to-light ratio becomes proportional
to 
\begin{eqnarray}\label{36}
\frac{M}{Nm}=k(r).
\end{eqnarray} 
Since in both the galactic bulge and outer region, $k(r)$ is an increasing function of the distance from the center of galaxy $r$ (see Eqs.(\ref{8}), (\ref{12})), hence we have much larger mass-to-light ratios than the central parts which is in complete agreement
with the Tully-Fisher relation.

From another point of view we may study the Tully-Fisher relation in the
present model. We may rewrite the equation (\ref{19}) as
\begin{eqnarray}\label{37}
v^4=\left(\frac{GNm}{R}\right)^2.
\end{eqnarray} 
If we consider $R$ as the characteristic size of the galaxy and $Nm$ as the
approximate stellar mass, then assuming a roughly constant mass density we
may write $Nm \sim R^3$ or $R^2$ according to whether we assume a spherical shape or disk shape for the whole galaxy, respectively. Therefore, Eq.(\ref{37}) reads
\begin{eqnarray}\label{38}
v^{\alpha} \sim Nm,
\end{eqnarray}
where $\alpha=3$ or 4 corresponding to spherical or disk shapes for the galaxy, respectively. This is an interesting result which again accounts
for the empirical Tully-Fisher relation.

\section{Liquid-Droplet model and MOND}

According to MOND, the rotational velocity is given by $v=(GNma_0)^{\frac{1}{4}}$.
However we find that in the present model this velocity is obtained by 
$v=\left(\frac{GNm}{R}\right)^{\frac{1}{2}}$. Comparison of these two velocities
leads to the following result
\begin{eqnarray}\label{39}
a_0=\frac{GNm}{R^2}.
\end{eqnarray}
This means that the constant acceleration $a_0$ which has no clear physical explanation in the MOND is explained in the present model through (\ref{39}) indicating that this acceleration is nothing but the {\it effective} gravitational strength of each structure. To justify this relation we consider for example the Andromeda galaxy whose stellar mass $Nm$ and bulge radius $R$ are approximately $7\times 10^{11} SM$ and $10^5 LY$, respectively. Putting these values into (\ref{39}) gives $a_0 \sim 10^{-10}$ which is in good agreement with the value obtained by Milgrom. It seems that similar calculation for other galaxies gives almost the same order of magnitude for the value of $a_0$. 

\section{Liquid-Droplet model and Dark matter}

The most direct assumption about the rotation curve problem is that there are a large number of nonluminous matters in the halo \cite{Halo}. There are lots of apparent evidences to show that dark matter exist, such as cosmic microwave background, dark matter in bullet cluster, and large-scale matter power spectrum \cite{Dark1}, however, as yet no direct evidences of the existence of dark matter have been found. This motivates one to search for other possible explanations for the rotation curve problem like MOND \cite{MOND}. The present Liquid-Droplet model, among other alternatives, shows its potential capability to solve the rotation curve problem by introducing a modification to the gravitational potential at large scale. Of course, it is not claimed that this model is better than dark matter or even MOND, however in our opinion if it can describe the observed rotation curve, at least it deserves to be considered as a potential alternative for dark matter or MOND. Indeed, regarding the matter component, the dark matter model introduces an extra and unknown component which can explain the rotation curves provided that we find its correct density distribution $\rho_{_{DM}}$(r). On the other hand, regarding the Newtonian dynamics, MOND introduces a modified Newtonian dynamics at very small accelerations which can explain the rotation curves provided that we find the correct forms of $\mu(x)$ or $g(r/r_0)$ defined in (\ref{mu(x)}) and (\ref{g(r)}), respectively. Similarly, in the present work, regarding the Newtonian gravity, we introduce a modified gravity at large scales which can explain the rotation curves provided that we find the correct form of $k(r)$ for each galaxy. Therefore, in each model we need to find some unknown properties like $\rho_{_{DM}}(r)$, $\mu(x)$ or $g(r/r_0)$, and $k(r)$. Contrary to dark matter and MOND models, where $\rho_{_{DM}}(r)$, $\mu(x)$ or $g(r/r_0)$ are partly characterized by some universal features like the universal behavior of dark matter or the universal acceleration $a_0$ introduced by MOND, the main advantage of the present model is that $k(r)$ is completely characterized
by the specific features of each individual galaxy and this feature may help
us to explain the diversity of rotation curves like flat, increasing, declining
and Keplerian and justify some observations as follows. 

\begin{itemize}
 \item It is known that few spirals galaxies exhibit decline in their rotation curves confirming the conventional belief that the mass distribution is truncated at about 1 to 3 optical radii (3-5 scale lengths) \cite{Cas}. In view of dark matter model, the decrease in rotation curve may be {\it interpreted} as an indication of a large ratio of luminous mass to dark matter in the luminous regions of these systems, a behaviour which is not in reasonable
agreement with the general expected behaviour of dark matter in the large scale structures. In view of the present model, however, the decrease in the rotation curve may be {\it explained} reasonably by an appropriate $k(r)$
given for instance by (\ref{8'}), (\ref{8''}),
 \item A detailed analysis shows a clear correlation between the specific features of a galaxy like peak circular velocity, its central surface brightness and the slope of the rotation curve in the outer components \cite{Cas}. This correlation may indicate a weakening of the well-known conspiracy between luminous and dark matter \cite{Cas}, and strengthen the present idea that the rotation curve is correlated with some specific features of each galaxy through $k(r)$. 
\item  Gravitational lensing studies of the Bullet cluster provides the best current evidence for the nature of dark matter \cite{Bullet}. In the Bullet cluster, a collision between two galaxy clusters appears to have caused a separation of dark matter and baryonic matter. X-ray observations show that much of the baryonic matter (gas) is concentrated in the center of the Bullet cluster. In fact, the passing gas particles are slowed down by electromagnetic interactions and settled near the point of impact. However, since dark matter components do not interact electromagnetically with each other, they are not slowed like the X-ray visible gas and are passed through each other, accompanied by galaxies, without slowing down substantially. This accounts for the separation of dark matter and baryonic matter. In view of the present model where there is no dark matter, we should consider the separation as a significant displacement between the center of visible matter and the center of gravitational potential. Contrary to the Newtonian potential, whose center is determined by the collective gravitational effect of all constituents in a gravitational system, the gravitational potential in our model is described by (\ref{3}) whose center is determined by the correlation function $k(r)$ and not the local position of constituents. We know that according
to Newtonian gravity the center of gravitational potential and center of visible mass should almost coincide in large structures. But, according to the present form of gravitational potential (\ref{3}) it is easy to see that there is no requirement for such coincidence. In fact, thanks to a
$k(r) \sim \alpha  r^n, 0<n<2$ the gravity force derived from (\ref{3}) at larger distances $r$ may be stronger. This feature may explain the gravitational lensing results on the Bullet cluster according to which we expect stronger gravitational force away from the cluster core. Moreover, for $k(r) \sim \alpha  r^n, -1<n<2$ the gravity force derived from (\ref{3}) at a given distance $r$ may be stronger than that of Newtonian gravity. This feature may also explain the weak gravitational lensing results on the galactic scales according to which we expect stronger gravitational force in the halo in
comparison to the Newtonian gravity. In fact, in this model, the necessity
for the existence of dark matter in producing enough strong gravity at larger distances to justify some observations like flat rotation curve or weak gravitational lensing, is replaced by the necessity for a modification in the gravitational potential in producing enough strong gravity at larger distances to justify the same observations.
\end{itemize}

\section{Conclusion}

The problem of rotation curves has been investigated in the
context of a new model which may be considered as an alternative to the dark
matter and MOND. This model is motivated by the so-called Weizs\"acker's
{\it Liquid-Droplet} model which had been used to describe the structure of the nucleus, according to which the nuclear force is saturated inside
the nucleus to exhibit a flat curve of binding energy. Similarly, we have shown that the flat rotation curve of galaxies or larger structures may have its origin in the saturation property of gravity at large scale structures. To establish such a property for gravity we have introduced a new model of gravitational potential by resorting to Machian viewpoint. This kind of gravitational potential at galactic or larger scales gives the rotation curve in good agreement with observations. The main advantage of our model lies in its flexibility which makes it possible to have one free function called ``correlation function'' $k(r)$ to adjust the theoretical predictions with observations, for any given structure. We have also developed the Virial theorem and also obtained the Tully-Fisher relation. Moreover, we have given a physical explanation for the so-called {\it constant} acceleration in the MOND, and also compared this model with Dark matter model.

Finally, it is worth noting that the results obtained for galactic structures are easily applicable for larger structures, like clusters of galaxies, with no loss of generality. This is because, the gravity is a long range force having scale invariance. Hence, if we replace the gravitational system
of a galaxy with the gravitational system of a cluster, all the arguments, reasonings and discussions at the galactic scales remain valid at the cluster scales. For instance, $k(r)$ becomes the correlation function between a mass $m$ (a galaxy) located at $r$, and the surrounding global mass of the cluster
$(Nm)$ whose center of mass is located at the origin $r=0$.

\section*{Acknowledgment}
I would like to thank the anonymous referee whose useful comments very much
improved the content of this manuscript. This work has been supported financially by Research Institute for Astronomy and Astrophysics of Maragha (RIAAM) under research project
NO.1/2782-3.

\end{document}